\documentclass[aps,pra,twocolumn,groupedaddress,showpacs,showkeys]{revtex4}

\usepackage{graphicx}
\begin{document}

\newcommand{\be}{\begin{equation}}
\newcommand{\ee}{\end{equation}}

\title{A scattering approach to Casimir forces and radiative heat transfer for nanostructured surfaces out of thermal equilibrium}

\author{Giuseppe Bimonte}
\email[Bimonte@na.infn.it]
\affiliation{Dipartimento di Scienze Fisiche Universit\`{a} di
Napoli Federico II Complesso Universitario MSA, Via Cintia
I-80126 Napoli Italy and INFN Sezione di Napoli, ITALY\\
}

\date{\today}

\begin{abstract}
We develop an exact method for computing  Casimir forces and the
power of radiative heat transfer between two arbitrary
nanostructured surfaces out of thermal equilibrium. The method is
based on a generalization   of the scattering approach recently
used in investigations on the Casimir effect. Analogously to the
equilibrium case, we find that also out of thermal equilibrium the
shape and composition of the surfaces enter only through their
scattering matrices. The expressions derived provide exact results
in terms of the scattering matrices of the intervening surfaces.
\end{abstract}


\maketitle

\section{Introduction}
\label{intro}

In recent years impressive technical advances have prompted
intensive experimental and theoretical investigations of proximity
phenomena, originating from quantum and thermal fluctuations of
the electromagnetic field existing in the vicinity of all bodies.
In general, these phenomena can be grouped in two broad classes,
namely equilibrium phenomena   one side, and non-equilibrium
phenomena on the other. Well-known examples of equilibrium
phenomena are the Casimir effect, and the Casimir-Polder atom-wall
forces \cite{parse,Mohid}. On the other hand, a much studied
non-equilibrium phenomenon is provided by non-contact radiative
heat transfer between two closely spaced bodies \cite{volokitin}.
In recent times much interest has been devoted to the new field of
Casimir and Casimir-Polder forces between macroscopic bodies
and/or atoms out of thermal equilibrium. These phenomena are
intensely investigated now both theoretically
\cite{antezza,buhmann} and experimentally \cite{cornell}. We
should also like to mention the very interesting phenomenon of
non-contact quantum friction \cite{pendry}. The problems of heat
transfer, Casimir forces and quantum friction in a system of two
plane-parallel plates at different temperatures, in relative
uniform motion in a direction parallel to the plates, have also
been investigated recently \cite{pers}.

While the material dependence of the above phenomena has been well
studied in simple planar geometries, there exists presently much
interest in exploring more complicated geometries. Indeed the
highly non-trivial geometry dependence of the near-field opens up
the possibility of tailoring the features of the radiation field
for new applications, that range from micro- and nanomachines
operated by the Casimir force \cite{chan,capasso}, to designing
the thermal emission of photonic crystals \cite{chan2}. For nearly
flat surfaces, the shape dependence can be studied using the
so-called proximity force approximation (PFA), which amounts to
averaging the plane-parallel result over the slowly varying
distance between the opposing surfaces of the bodies. It is well
known however that the PFA has only a limited range of validity,
and it can lead to inaccurate predictions for surfaces with deep
corrugations \cite{chan3}. For this reason, it is widely
recognized today that more accurate methods are needed to describe
arbitrary geometries. For systems in equilibrium, much progress
has been made recently. New powerful numerical techniques to
compute the Casimir force,  based on the Green's function approach
and on the path-integral approach have been reported
\cite{gies,rodriguez}. Another approach that is being vigorously
developed is based on the multiple scattering formalism, that was
first introduced long ago \cite{balian} to study the Casimir
energy for a system of perfect conductors of arbitrary shape.
The scattering approach is actually closely related to the Green's
function method, and we address the reader to Ref. \cite{balian}
for further details on the connection between the two methods.  
There exist today several variants of the scattering approach,  that have been
developed to deal also with real materials (for a comparative
review see \cite{milton}): one of the variants \cite{emig,kenneth}
is better suited for dealing with compact objects not too close to
each other, and it is based on a multipolar expansion of the e.m.
field. Another variant \cite{genet} is instead better adapted to
deal with planar-like structures in close proximity, and it uses a
decomposition of the e.m. field  into plane waves. Contrasted with
these important advances in the mathematical techniques for
computing equilibrium Casimir forces, the theory is much less
developed for problems out of thermal equilibrium. In fact, the
Casimir force out of thermal equilibrium has been investigated
only in the plane-parallel case \cite{antezza}, while the problem
of computing the near-field radiative transfer between two spheres
was addressed only very recently \cite{chen}, using a dyadic
Green's function approach (see also the recent experiment
\cite{nara}). Investigating in depth the shape-dependence of
proximity effects out of thermal equilibrium is indeed very
interesting, in view of potential applications, because out of
equilibrium there exists a richness of behaviors, associated for
example with resonances in the spectrum of surface excitations,
that are absent at equilibrium \cite{volokitin,antezza}.

In this  paper we develop an exact method for computing Casimir
forces and the power of heat transfer between two arbitrary plates
out of thermal equilibrium. The method is based on a
generalization  of the scattering approach, that has proven so
successful in equilibrium Casimir problems. We consider the
variant of the scattering approach \cite{genet} that is best
suited for planar-like nanostructured surfaces at close
separations, like those of \cite{chan2,chan3}.  In this paper, we
shall only present the derivation of the basic formulae, leaving
concrete numerical applications for a successive exposition. The
main new result is the demonstration that also out of thermal
equilibrium, the shape and material dependence  enter only through
the scattering matrices of the bodies involved, analogously to
what has been found for systems in thermal equilibrium. We remark
that our results provide exact expressions in terms of the
scattering matrices of the intervening bodies. Of course, the
scattering matrix is in principle a complicated object, but there
exist methods, both analytical and numerical, for computing it
accurately. Many new geometries have been considered recently in
Casimir investigations, and the corresponding scattering matrices
have been estimated for this purpose. The formulae derived in this
paper permit to consider these geometries out of thermal
equilibrium.

The plan of the paper is as follows: in Sec. 2 we present the
general derivation of the correlators for the e.m. field in the
gap between two arbitrarily shaped plates at different
temperatures. In Sec. 3 the correlators derived in Sec. 2 are used
to obtain an exact expression for the Casimir force and the power
of radiative heat transfer between the plates. Finally, in Sec. 4
we present our conclusions and outline directions for future work.

\section{General principles: Rytov's theory.}

We consider a  geometry of the type considered in Refs.
\cite{genet}, i.e. a cavity consisting of two large plates plates
at temperatures $T_1$ and $T_2$, the whole system being in a {\it
stationary configuration}. The shapes of their opposing surfaces
can be arbitrary, apart from the assumption (usually implicit in
scattering approaches to equilibrium Casimir problems
\cite{genet,kenneth,emig}) that there must exist between them a
vacuum gap of thickness $a>0$ bounded by two parallel planes. This
condition excludes from our consideration interpenetrating
surfaces, like the one studied in the second of Refs.
\cite{rodriguez}. We also assume for simplicity that the plates
are thick, in such a way that no radiation from outside can enter
the gap.

The basic problem that we face   is to determine the correlators
for the fluctuating electromagnetic field existing in the empty
gap between the plates. This can be done by suitably generalizing
the methods used in heat transfer studies \cite{volokitin}, which
are also at the basis of the recent out-of-equilibrium Casimir
investigations \cite{antezza}. Both are based on the well known
Rytov's theory \cite{rytov} of electromagnetic fluctuations. On
the basis of this theory, the field in the gap can be interpreted
as the result of multiple scatterings off the plates surfaces, of
the radiation fields originating from   quantum and thermal
fluctuating polarizations within the plates. Importantly, the
local character of the polarization fluctuations implies that the
two plates radiate {\it independently} from each other. Therefore,
in a {\it stationary configuration}, the radiation from either
plate is the {\it same} as the one   that would be radiated by
that plate, if it were in {\it equilibrium} with the environment
(at its own temperature), the other plate being {\it removed}.
This physical picture permits to separate the problem of
determining the fluctuating field in the gap in two separate
steps. In the first step, one determines the field radiated by
either plate in isolation, a problem that can be solved by using
the general equilibrium formalism. The two plates are considered
together only in the second step, where the intracavity field is
finally determined, by taking account of the effect of multiple
scatterings on the radiation fields radiated by the plates, as
found in step one.

After these general remarks, we can now start our two-step
computation of the fluctuating e.m. field in the gap. We let
$\{x,y,z\}$ cartesian coordinates such that the vacuum gap is
bounded by the planes   $z=0$ and $z=a$ respectively, with plate
one (two) lying at the left (right) of the $z=0$ ($z=a$) plane. We
suppose that the lateral sizes $L_x$ and $L_y$ of the plates are
both much larger than the separation $a$: $L_x,L_y \gg a$ Boundary
effects being negligible, it is mathematically convenient to
impose periodic boundary conditions on the fields in the $(x,y)$
directions, on the opposite sides of the plates. When dealing with
the e.m. field, it is sufficient to consider the electric field
${\bf E}(t,{\bf r})$ only, for the magnetic field ${\bf B}(t,{\bf
r})$ can be obtained from ${\bf E}(t,{\bf r})$ by using Maxwell
Equations. The geometry being planar-like, the electric field in
the gap can always be expressed as a sum of (positive-frequency)
plane-wave modes of the form \be {\bf E}^{(\pm)}_{\alpha,{\bf
k}_{\perp} }(t,{\bf r})=2\,{\rm Re}\,[b^{(\pm)}_{\alpha,{\bf
k}_{\perp} }{\bf \cal E}^{(\pm)}_{\alpha,{\bf k}_{\perp}
}({\omega;\bf r}) \,e^{-i \omega t}]\label{mode0}\ee where \be
{\bf \cal E}^{(\pm)}_{\alpha,{\bf k}_{\perp} }({\omega;\bf
r})={\bf e}^{(\pm)}_{\alpha,{\bf k}_{\perp} }(\omega)\,e^{i {\bf
k}^{(\pm)}\cdot {\bf r}}\;. \label{modes}\ee Here $\omega$ is the
frequency, and ${\bf k}_{\perp}$ is the projection of the
wave-vector onto the $(x,y)$ plane. Periodicity in $(x,y)$
directions implies that the wave-vectors ${\bf k}_{\perp}$ belong
to a discrete set labelled by two integers $(n_x,n_y)$: $k_x=2 \pi
n_x/L_x$, $k_y=2 \pi n_y/L_y$. The index $\alpha=s,p$ denotes
polarization, where $s$ and $p$ correspond, respectively, to
transverse electric and transverse magnetic polarizations. The
superscripts (+) and (-) in Eqs. (\ref{mode0}) and (\ref{modes})
refer to the direction of propagation along the $z$-axis, the
$(+)$ and $(-)$ signs corresponding to propagation in the positive
and negative $z$ directions, respectively. Moreover, ${\bf
k}^{(\pm)}={\bf k}_{\perp} \pm k_z {\hat {\bf z}}$, where
$k_z=\sqrt{\omega^2/c^2-k^2_{\perp}}$ (the square root is defined
such that ${\rm Re}(k_z) \ge 0$, ${\rm Im}(k_z) \ge 0$), ${\bf
e}^{(\pm)}_{s,{\bf k}_{\perp}}(\omega)={\hat {\bf z}} \times {\hat
{\bf k}}_{\perp}$, ${\bf e}^{(\pm)}_{p,{\bf
k}_{\perp}}(\omega)=(c/\omega)\,{\bf k}^{(\pm)} \times {\bf
e}^{(\pm)}_{s,{\bf k}_{\perp}}$. We note that for $\omega/c
> k_{\perp}$, when $k_z$ is real, the modes ${\bf \cal
E}^{(\pm)}_{\alpha,{\bf k}_{\perp} }$ represent {\it propagating}
waves, while for $\omega/c < k_{\perp}$, when $k_z$ is imaginary,
they describe {\it evanescent} modes.  It is opportune to
introduce a shortened index notation, that will prove useful in
the sequel. We shall use a roman index $i$ to denote collectively
the $i$-th component of a vector and the position ${\bf r}$, while
a greek index $\alpha$ will denote collectively the polarization
$\alpha$ and the wave-vector ${\bf k}_{\perp}$. In this notation
the $i$-th component of   ${\bf {\cal E}}^{(\pm)}_{\alpha,{\bf
k}_{\perp} }(\omega;{\bf r})$ shall be denoted as ${\cal
E}^{(\pm)}_{i \alpha}(\omega)$. Similarly, a kernel
$A_{\alpha,{\bf k}_{\perp};\alpha',{\bf k}'_{\perp}}(\omega)$
shall be denoted  as $A_{\alpha,\alpha'}(\omega)$. We also set
$\sum_\omega \equiv \int d \omega/(2 \pi)$, $\delta_{\omega,
\omega'}\equiv 2 \pi \delta(\omega-\omega')$, $\sum_{\alpha}
\equiv  1/{\cal A} \sum_{n_x,n_y}\sum_{\alpha}$, and
$\delta_{\alpha,\alpha'}\equiv {\cal A}\, \delta_{n_x,n'_x}
\delta_{n_y,n'_y}\delta_{\alpha,\alpha'}$, where ${\cal A}=L_x
L_y$ is the area of the plates. Finally, for any kernel
$A_{\alpha,{\bf k}_{\perp};\,\alpha',{\bf k}'_{\perp}}$, we define
$${\rm Tr}_{\alpha} A= \sum_{\alpha} A_{\alpha,\alpha}\equiv
\frac{1}{\cal A}\sum_{n_x,n_y}\sum_{\alpha} A_{\alpha,{\bf
k}_{\perp};\,\alpha,{\bf k}_{\perp}}\;.$$ Having set our
notations, we  pass now to step one.

\subsection{Step one: the field radiated by a single plate in thermal equilibrium}

As explained above, we begin by considering each plate in
isolation to determine its radiation, and we let ${{\cal
E}}^{(A)}_i(\omega)\;,A=1,2$ the time Fourier-transform of the
field radiated by plate $A$. The {\it total} radiation field
${{\cal E}}^{({\rm eq};A)}_i(\omega )$ existing, respectively, to
the right of plate one and to the left of plate two, when either
plate is in equilibrium at temperature $T_A$ with the environment
(the other plate being absent) can be expressed in the form: \be
{{\cal E}}^{({\rm eq};A)}_i(\omega)={{\cal E}}^{(A)}_i(\omega)+{
\cal E}^{({\rm env};A)}_i(\omega)+{\cal E}^{(\rm
sc;A)}_i(\omega)\;,\label{etot}\ee where ${ \cal E}^{({\rm
env};A)}_i(\omega)$ describes the environment radiation, including
vacuum fluctuations and black-body radiation, impinging on plate
$A$ (from the right for plate one, and from the left for plate
two), and ${\cal E}^{({\rm sc},A)}_i(\omega))$ is the
corresponding scattered radiation. These fields have the
expansions: \be {{\cal E}}^{(A)}_i(\omega)=\sum_{\alpha}
  {\cal E}^{(\pm)}_{i \alpha}(\omega)\,b^{(A)}_{\alpha}(\omega)\,\;\label{Afield}\;,\ee
  \be {{\cal
E}}^{({\rm env};A)}_i(\omega)=\sum_{\alpha}
  {\cal E}^{(\mp)}_{i \alpha}(\omega)\,b^{(\rm env)}_{\alpha}(\omega)\,\;\label{bbfield}\;,\ee
   \be {{\cal
E}}^{({\rm sc};A)}_i(\omega)=\sum_{\alpha,\alpha'} {\cal
E}^{(\pm)}_{i
\alpha}(\omega)\,S^{(A)}_{\alpha,\alpha'}(\omega)\,b^{(\rm
env)}_{\alpha'}(\omega)\,,\label{scfield}\ee where, here and in
Eqs. (\ref{scapos}) and (\ref{green}) below, the upper (lower)
sign is for plate one (two), and $S_{\alpha \alpha'}^{(A)}$ is the
scattering matrix of plate $A$, for radiation impinging on the
right (left) surface of plate one (two). It is important to note
that, for fixed plates orientations, the matrix $S_{\alpha
\alpha'}^{(A)}$ depends in general on the position ${\bf x}^{(A)}$
of some fixed reference point $Q^{(A)}$ chosen on plate $A$. If
${\tilde S}_{\alpha \alpha'}^{(A)}$ is the scattering matrix of
plate $A$ relative to a coordinate system with origin at
$Q^{(A)}$, then: \be {S}^{(A)}_{\alpha \alpha'}=e^{-i {\bf
k}^{(\pm)}\cdot {\bf x}^{(A)}}\,{\tilde S}^{(A)}_{\alpha
\alpha'}\, e^{i {\bf k}^{'(\mp)}\cdot {\bf
x}^{(A)}}\;.\label{scapos}\ee The amplitudes $b^{(\rm
env)}_{\alpha}(\omega)$ for the environment radiation in Eqs.
(\ref{bbfield}) and (\ref{scfield}) are characterized by the
following non-vanishing well known correlators: \be \langle
b^{(\rm env)}_{\alpha}(\omega)\, b^{(\rm env)*}_{\alpha'}(\omega')
\rangle= \frac{2 \pi \omega}{c^2} \,F(\omega, T_A)\, {\rm
Re}\left( \frac{1}{k_z}\right)\delta_{\omega \omega'}
\delta_{\alpha \alpha'}\,\label{bbcor}\ee where
$F(\omega,T)=(\hbar \omega/2) \coth(\hbar \omega/(2 k_B T))$, with
$k_B$ Boltzmann constant.
The desired correlators for the amplitudes
$b^{(A)}_{\alpha}(\omega)$ can now be determined by exploiting the
following   relation implied by the fluctuation-dissipation
theorem: \be \langle {{\cal E}}^{({\rm eq}; A)}_i(\omega )\,{{\cal
E}}^{({\rm
eq};A)*}_{i'}(\omega')\rangle=\frac{2}{\omega}\,F(\omega,T_A)\,
\delta_{\omega \omega'}{\rm Im} \,
G^{(A)}_{ii'}(\omega)\;,\label{FDT}\ee where
$G^{(A)}_{ii'}(\omega)$ is the dyadic $retarded$ Green function of plate $A$.
In the vacuum to the right (left) of plate one (two), the Green
function $G^{(A)}_{ii'}(\omega)$ can be expressed in terms of the
scattering matrix $S^{(A)}_{\alpha \alpha'}$ as follows: \be
G^{(A)}_{ii'}(\omega)=G_{ii'}^{(0)}(\omega)+ \frac{2 \pi i
\omega^2}{c^2} \sum_{\alpha \alpha'}  {\cal
E}_{i\alpha}^{(\pm)}S_{\alpha \alpha'}^{(A)} {\cal
E}_{J(i')\alpha'}^{(\mp)}\frac{1}{k'_z}\;,\label{green}\ee where
$G_{ii'}^{(0)}(\omega)$ is the $retarded$ Green function in free space:
$$
G_{ii'}^{(0)}(\omega)=\frac{2 \pi i \omega^2}{c^2}
\sum_{\alpha}\frac{1}{k_z}\left(\theta(z-z') \,{\cal
E}_{i\alpha}^{(+)} {\cal E}_{J(i')\alpha}^{(+)}\right.$$\be\left.+
\,\theta(z'-z)\, {\cal E}_{i\alpha}^{(-)} {\cal
E}_{J(i')\alpha}^{(-)}\right)\;,\label{freegr}\ee with $\theta(z)$
Heaviside step-function ($\theta(z)=1$ for $z\ge 0$, $\theta(z)=0$
for $z<0$). Here, $J$ denotes the {\it inversion} operator, whose
action on space-indices is defined as $J(i) \equiv J(i,{\bf
r})=(i,-{\bf r})$. It is useful to define the action of $J$ also
on polarizations, wave-vectors and propagation directions as
$J(\alpha)\equiv J(\alpha,{\bf k}_{\perp})=(\alpha,-{\bf
k}_{\perp})$ and $J((\pm))=(\mp)$. The following relations hold
\be {\cal E}^{(\pm) *}_{i \alpha}={\cal E}^{(\pm)}_{J(i)
\alpha}(1+s_{\alpha})/2 +{\cal E}^{(\mp)}_{J(i)
\alpha}(1-s_{\alpha})/2 \;,\label{rel1}\ee where $s_{\alpha}
\equiv {\rm sign}(\omega^2/c^2-k^2_{\perp})$ and \be {\cal
E}^{(\pm)}_{J(i) J(\alpha)}=(-1)^{P(\alpha)}{\cal
E}^{(\mp)}_{ia}\;,\label{rel2}\ee where $P(\alpha)$ is one (zero)
for $s$ ($p$) polarization.  The reciprocity relations
$G^{(A)}_{ii'}(\omega)=G^{(A)}_{i'i}(\omega)$ satisfied by the
Green's function, as a consequence of microscopic reversibility,
imply via Eq. (\ref{green}) the following important Onsager's
relations that must hold for any scattering matrix \be
S^{(A)}_{\alpha
\alpha'}=\frac{k'_z}{k_z}(-1)^{P(\alpha)+P(\alpha')}S^{(A)}_{J(\alpha')J(\alpha)}\;.\label{ons}\ee
Upon substituting the expression for ${\cal E}^{({\rm
eq};A)}_i(\omega )$ provided by Eqs.(\ref{etot}-\ref{scfield})
into the l.h.s. of Eq. (\ref{FDT}), and after substituting the
expression of the Green function Eqs. (\ref{green}) into the
r.h.s. of Eq. (\ref{FDT}), by making use of Eqs. (\ref{bbcor}),
(\ref{rel1}), (\ref{rel2}) and (\ref{ons}) one obtains the
following expression for the non-vanishing correlators of the
amplitudes $b^{(A)}_{\alpha}(\omega)$: $$ \langle b^{(A)}
(\omega)\, b^{(B)\dagger} (\omega') \rangle =\delta_{AB}\,\frac{2
\pi \omega}{c^2} F(\omega, T_A)\, \delta_{\omega \omega'}$$ \be
\times \left(\Sigma^{\rm (pw)}_{-1} - S^{(A)} \Sigma^{\rm
(pw)}_{-1} S^{(A)\dagger}+S^{(A)} \Sigma^{\rm
(ew)}_{-1}-\Sigma^{\rm (ew)}_{-1} S^{(A)\dagger}
\right)\,,\label{kirgen}\ee where we collected the amplitudes
$b^{(A)}_{\alpha}(\omega)$ into the (column) vector
$b^{(A)}(\omega)$ and we set $\Sigma_{n}^{\rm (pw/ew)}=k_z^n
\Pi^{\rm (pw/ew)} $, where $\Pi^{\rm (pw)}_{\alpha \alpha'}=
\delta_{\alpha \alpha'}\,{(1+s_{\alpha})}/{2}$ and $\Pi^{\rm
(ew)}_{\alpha \alpha'}= \delta_{\alpha
\alpha'}\,{(1-s_{\alpha})}/{2}$ are the projectors onto the
propagating and evanescent sectors, respectively. Eq.
(\ref{kirgen}) generalizes the well known Kirchhoff's law (as can
be found for example in \cite{volokitin}) to non-planar surfaces,
and it shows that the fluctuating field radiated by plate $A$ is
fully determined by its scattering matrix $S^{(A)}$. We remark
that for non-planar surfaces the matrix $S^{(A)}$ is non-diagonal,
and therefore the order of the factors on the r.h.s. of Eq.
(\ref{kirgen}) must be carefully respected. Now  we move to step
two.

\subsection{Step two: determination of the intracavity field}

Without loss of generality, the   intra-cavity field  can be
represented as a superposition of waves of  the form: \be {\cal
E}_{i \alpha}(\omega)=b_{\alpha}^{(+)}(\omega){\cal
E}_{i\alpha}^{(+)}(\omega)+b_{\alpha}^{(-)}(\omega){\cal
E}_{i\alpha}^{(-)}(\omega)\;.\ee The intuitive physical picture of
the intra-cavity field as resulting from repeated scattering off
the two surfaces of the radiation field {\it emitted} by the
surfaces of the two plates leads  to the following equations for
  $b^{(\pm)}(\omega)$:
\be b^{(+)}=b^{(1)}+{S}^{(1)}\,b^{(-)}\;,\;\;
b^{(-)}=b^{(2)}+{S}^{(2)}\,b^{(+)}\;.\label{intrafield}\ee
 Equations (\ref{intrafield})  are easily solved: \be b^{(+)}= U^{(12)}\,b^{(1)}+{S}^{(1)} U^{(21)}\,
 b^{(2)}\;,\label{bplus}\ee \be
b^{(-)}={S}^{(2)} U^{(12)}\, b^{(1)}+  U^{(21)}\, b^{(2)}
\;,\label{bminus}\ee where $U^{(AB)}=(1-{
S}^{(A)}{S}^{(B)})^{-1}$. Together with Eq. (\ref{kirgen}), Eqs.
(\ref{bplus}) and (\ref{bminus}) completely determine the
intra-cavity field. In particular, they determine the matrix
$C^{(KK')}$ for the non-vanishing correlators of the intracavity
field: \be\langle b^{(K)}(\omega) b^{(K')\dagger}(\omega')
\rangle= \delta_{\omega, \omega'}C^{(KK')}\;.\ee The explicit
expression of $C^{(KK')}$ in terms of ${S}^{(1)}$ and ${S}^{(2)}$
can be easily obtained from Eqs. (\ref{kirgen}), (\ref{bplus}) and
(\ref{bminus}), and it is not shown for brevity.

\section{Observables}

The above results permit   to evaluate the average of any
observables constructed out of the intracavity field. Typically,
the observables are symmetric bilinears of the electric field, of
the form \be {\bar {\cal O}} \equiv \sum_{i j}\int d^2{\bf
r}_{\perp} \int d^2 {\bf r}'_{\perp} E_i(t,{\bf r}) {\cal
O}_{ij}({\bf r},{\bf r}')E_j(t,{\bf r}')\;,\ee where ${\cal
O}_{ij}({\bf r},{\bf r}')={\cal O}_{ji}({\bf r'},{\bf r})$.  Upon
defining the matrix $$ {{\cal O}}^{(KK')}_{\alpha,\alpha'}=\sum_{i
j}\int d^2{\bf r}_{\perp}\!\! \int d^2 {\bf r}'_{\perp} {\cal
E}_{i\alpha}^{(K)*}(\omega,{\bf r})$$ \be \times {\cal
O}_{ij}({\bf r},{\bf r}')\,{\cal E}_{j\alpha'}^{(K')}(\omega,{\bf
r}')\ee the statistical average of ${\bar {\cal O}}$ can be
written as
 \be\langle {\bar {\cal O}}\rangle= 2 \sum_{\omega >0}
\sum_{K,K'} {\rm Tr}_{\alpha} [C^{(KK')} {{\cal O}}^{(K'K)}
]\,.\label{aver}\ee Below we shall use this formula to determine
the Casimir force and the power of heat transfer between the two
plates.

\subsection{The Casimir force out of thermal equilibrium}

As our first example, we consider the $(x,y)$ integral of the $zz$
components  of the Maxwell stress tensor $T_{ij}$, that provides
the total Casimir force between the plates.  After a simple
computation, one finds: \be { {\cal
O}}^{(KK')}\,[T_{zz}]=\frac{c^2 k^2_z}{4 \pi \omega^2}\, \left(
\delta_{KK'}\,\Pi^{(\rm pw)}+ \delta_{KJ(K')}\,\Pi^{(\rm
ew)}\right)\;.\label{Tzz}\ee Evaluation of Eq. (\ref{aver}) with
${\cal O}^{(KK')}$ given by Eq. (\ref{Tzz}), leads to the
following representation for the {\it unrenormalized} Casimir
force $F_z^{(0\,\rm neq)}$  out of thermal equilibrium: $$
F_z^{(0\,\rm neq)} \!=\!\sum_{\omega
>0}\frac{1}{\omega}\,[F(\omega,T_1)J(S^{(1)},S^{(2)}) $$ \be   + \;F(\omega,T_2)J(S^{(2)},S^{(1)})]\;,\label{unren}\ee
where $J(S^{(A)},S^{(B)})$ is the quantity $$ J(S^{(A)},S^{(B)})=
 {\rm Tr_{\alpha}}\! \left[
U^{(AB)}\left(\Sigma^{\rm (pw)}_{-1} -S^{(A)}\Sigma^{\rm
(pw)}_{-1}S^{(A)\dagger}\right. \right.
$$
$$
\left.+S^{(A)}\Sigma^{\rm (ew)}_{-1}-\Sigma^{\rm
(ew)}_{-1}S^{(A)\dagger}\right) U^{(AB)\dagger}\,\left(
\Sigma^{\rm (pw)}_{2}\right.$$ \be \left.\left. + S^{(B)\dagger}
\Sigma^{\rm (pw)}_{2}S^{(B)}+ \Sigma^{\rm (ew)}_{2}
S^{(B)}+S^{(B)\dagger} \Sigma^{\rm (ew)}_{2} \right)\right]
\;.\label{jqua}\ee After we add and subtract one half of the
quantity
$$B=F(\omega,T_2)J(S^{(1)},S^{(2)})+
F(\omega,T_1)J(S^{(2)},S^{(1)})$$ from the expression inside the
square brackets on the r.h.s. of Eq. (\ref{unren}), it is easily
seen that Eq. (\ref{unren}) can be recast in the form: $$
F_z^{(0\,\rm neq)}(T_1,T_2) =\frac{F_z^{(0\,\rm
eq)}(T_1)+F_z^{(0\,\rm eq)}(T_2)}{2}$$ \be +\;\Delta F_z^{(\rm
neq)}(T_1,T_2)\;,\label{unrbis}\ee where \be F_z^{(0\,\rm eq)}
\!=\!\sum_{\omega
>0}\frac{1}{\omega}F(\omega,T)[J(S^{(1)},S^{(2)})+J(S^{(2)},S^{(1)})]\;,\label{unreq} \ee
and $$\Delta F_z^{(\rm neq)}(T_1,T_2)= \sum_{\omega
>0}\frac{1}{2 \omega}(F(\omega,T_1)-F(\omega,T_2))$$
\be \times
\,[J(S^{(1)},S^{(2)})-J(S^{(2)},S^{(1)})]\;.\label{neqf0}\ee Using
the identity \be F(\omega,T)=\hbar
\omega\left[\frac{1}{2}+n(\omega,T)\right]\ee where \be
n(\omega,T)=\frac{1}{\exp(\hbar \omega/(k_B T))-1}\;,\ee Eq.
(\ref{neqf0}) can be written as:
$$\Delta F_z^{(\rm neq)}(T_1,T_2)= \frac{\hbar}{2}\sum_{\omega
>0}(n(\omega,T_1)-n(\omega,T_2))$$
\be \times
\,[J(S^{(1)},S^{(2)})-J(S^{(2)},S^{(1)})]\;.\label{neqf}\ee On the
 other hand, upon substituting Eq. (\ref{jqua}) into the r.h.s. of Eq.
(\ref{unreq}), after a somewhat lengthy algebraic manipulation, it
can be seen that the quantity $F_z^{(0\,\rm eq)}(T)$ can be
further decomposed as  \be F_z^{(0\,\rm eq)}(T)=A^{(0)}(T)+
F_z^{(\rm eq)}(T)\;.\label{unreqbis}\ee Here, $A^{(0)}(T)$ denotes
the divergent quantity: \be A^{(0)}(T)=2\sum_{\omega
>0}  \frac{F(\omega,T)}{\omega} {\rm
Tr_{\alpha}}\left[k_z \Pi^{\rm (pw)}\right].\ee As we see, this
quantity depends neither on the material constituting the plates
nor on their distance, and we neglect it altogether \footnote{In
effect, the divergent quantity $A^{(0)}(T)$ includes a {\it
finite} temperature-dependent contribution, which may give rise to
a distance-independent force on the plates. The actual magnitude
of the resulting constant force on either plate depends on the
temperature of the environment outside the cavity, but it is
independent of both the material constituting the plates, as well
as of their shapes. For a detailed discussion of this point, the
reader may consult the third of Refs.\cite{antezza}}. As to the
second contribution $F_z^{(\rm eq)}$ occurring on the r.h.s. of
Eq. (\ref{unreqbis}), it has the expression
$$  F_z^{(\rm eq)}(T)=2 \,{\rm Re} \sum_{\omega \ge 0}
\frac{F(\omega,T)}{\omega} \,{\rm Tr}_{\alpha}\,\left[k_z\left(
U^{(12)}\,S^{(1)} S^{(2)}\right. \right.$$ \be \left. \left.+
\,U^{(21)}\,{S^{(2)} S^{(1)}}  \right) \right]\;\label{eqcas}.\ee
Recalling that, according to Eq. (\ref{scapos}), the scattering
matrices $S^{(A)}$ depend on the mutual positions of the plates,
it is easy to verify that the above equilibrium force $F_z^{(\rm
eq)}$ has an associated free energy $F(a,T)$ ($F_z^{(\rm
eq)}=\partial F(a,T)/\partial a$) equal to: \be F(a,T)=2 \,{\rm
Im} \sum_{\omega \ge 0} \frac{F(\omega,T)}{\omega} \,{\rm
Tr}_{\alpha} \log (1-S^{(1)} S^{(2)})\;.\label{freen}\ee  Eqs.
(\ref{eqcas}) and (\ref{freen}) coincide with the equilibrium
expressions, as derived within the scattering approach
\cite{genet}. Putting everything together, after in Eq.
(\ref{unrbis}) we remove the divergent contribution proportional
to $A^{(0)}(T_1)+A^{(0)}(T_2)$, we obtain the following new {\it
exact} expression for the renormalized Casimir force between the
plates: \be F_z^{(\rm neq)}(T_1,T_2)=\frac{ F_z^{(\rm
eq)}(T_1)+F_z^{(\rm eq)}(T_2)}{2}+\,\Delta F_z^{(\rm
neq)}(T_1,T_2)\;.\label{renneq}\ee Some comments are now in order.
We note first of all that the quantities $ F_z^{(\rm eq)}(T)$ and
$\Delta F_z^{(\rm neq)}(T_1,T_2)$ are both {\it finite}. Indeed,
as we said earlier,   our expression for $F_z^{(\rm eq)}(T)$
coincides with the known scattering-approach expression for the
equilibrium Casimir force, which has been shown to be finite in
previous studies \cite{genet}. As to $\Delta F_z^{(\rm
neq)}(T_1,T_2)$, it is apparent from Eq. (\ref{neqf}) that this
quantity is  finite, thanks to the Boltzmann factors
$n(\omega,T_i)$. We also note that our result has the same general
structure as the formula derived in Refs.\cite{antezza}, for the
simpler case of two plane-parallel plates. Analogously to that
case, we indeed see from Eq. (\ref{renneq}) that the
non-equilibrium force is the sum of the average of the equilibrium
forces, for the temperatures $T_1$ and $T_2$, plus a contribution
$\Delta F_z^{(\rm neq)}(T_1,T_2)$, that vanishes for $T_1=T_2$
(see Eq. (\ref{neqf})).   Moreover, it is interesting to observe
that even for $T_1 \neq T_2$, the quantity $\Delta F_z^{(\rm
neq)}$, being antisymmetric in the  scattering matrices of the two
plates, vanishes if the two plates have identical scattering
matrices. Such a case is realized, for example, if the two plates
are made of the same material and if their profiles are specularly
symmetric with respect to the $(x,y)$ plane. When this happens,
the non-equilibrium Casimir force is just the average of the
equilibrium forces, for  the two temperatures of the plates. An
analogous statement  can be found in the third of Refs.
\cite{antezza}.  We can easily verify that in the case of
plane-parallel homogeneous dielectric plates our general formula
Eq. (\ref{renneq}) reproduces the result of Refs. \cite{antezza}.
In the flat case the scattering matrices of the plates are
diagonal, and can be taken to be of the form
\begin{eqnarray}
  S^{(1)}_{\alpha
\alpha'} &=& \delta_{\alpha \alpha'} R_{\alpha}^{(1)}\;,\nonumber\\
  S^{(2)}_{\alpha \alpha'} &=& \delta_{\alpha \alpha'} R_{\alpha}^{(2)}\,e^{2 i k_z
  a}\;,\label{plane}
\end{eqnarray}
where $R_{\alpha}^{(A)}$ denote the familiar Fresnel reflection
coefficients. When these diagonal scattering matrices are plugged
into Eq. (\ref{eqcas}), one obtains:  $$  F_z^{(\rm
eq)}(T)=4\,{\cal A}\,{\rm Re} \sum_{\omega \ge 0}
\frac{F(\omega,T)}{\omega} \sum_{\alpha} k_z
\frac{R_{\alpha}^{(1)}\,R_{\alpha}^{(2)}\,e^{2 i k_z
a}}{1-R_{\alpha}^{(1)}\,R_{\alpha}^{(2)}\,e^{2 i k_z
  a}}$$
  \be
 =4
{\cal A} \,{\rm Re} \sum_{\omega \ge 0} \frac{F(\omega,T)}{\omega}
\sum_{\alpha} k_z \left[\frac{e^{-2 i k_z
a}}{R_{\alpha}^{(1)}\,R_{\alpha}^{(2)}} -1\right]^{-1}\;.
\label{lifs}\ee In the limit of infinite plates, when
$$\frac{1}{\cal A}\sum_{n_x,n_y}\rightarrow \int \frac{d^2{\bf k}_{\perp}}{(2
\pi)^2}\;,$$ the above formula reproduces the well known Lifshitz
formula \cite{parse} for the Casimir force between two dielectric
plane-parallel slabs. On the other hand, when the scattering
matrices in Eq. (\ref{plane}) are substituted into Eq.
(\ref{neqf}), one finds:
$$\Delta F_z^{(\rm neq)}(T_1,T_2)={\cal A} \times {\hbar}\sum_{\omega
>0}[n(\omega,T_1)-n(\omega,T_2)]$$ $$\times \sum_{\alpha}\left[{\rm Re}(k_z)
\frac{|R_{\alpha}^{(2)}|^2-|R_{\alpha}^{(1)}|^2
}{|1-R_{\alpha}^{(1)}\,R_{\alpha}^{(2)}\,e^{2 i k_z
  a}|^2}-2 \,{\rm Im}(k_z)\,e^{-2 a{\rm Im}(k_z)}\right.$$
  \be\left.
  \times \frac{{\rm Im}(R_{\alpha}^{(1)}){\rm Re}(R_{\alpha}^{(2)})-
  {\rm Re}(R_{\alpha}^{(1)}){\rm Im}(R_{\alpha}^{(2)})}{|1-R_{\alpha}^{(1)}\,R_{\alpha}^{(2)}\,e^{2 i k_z
  a}|^2}\right]\;.\label{delplane}\ee
After we substitute Eqs. (\ref{lifs}) and (\ref{delplane}) into
Eq. (\ref{renneq}), and upon taking the limit of infinite plates,
one finds that the result coincides with the non-equilibrium
Casimir force computed in Refs.\cite{antezza}.

\subsection{Power of heat transfer}

We consider now the total power $W$ of heat transfer between the
plates. This requires  that we evaluate the statistical average of
the $(x,y)$ integral of the $z$-component $S_z$ of the Poynting
vector in the gap between the plates. A simple computation shows
that: \be { {\cal O}}^{(KK')}\,[S_{z}]=\frac{c^2 k_z}{4 \pi
\omega}\,(-1)^K \left( \delta_{KK'}\,\Pi^{(\rm pw)}+
\delta_{KJ(K')}\,\Pi^{(\rm ew)}\right)\;.\label{Sz}\ee When this
expression is plugged into Eq. (\ref{aver}) we obtain: \be W
=\sum_{\omega
>0}[F(\omega,T_1)H(S^{(1)},S^{(2)})-F(\omega,T_2)H(S^{(2)},S^{(1)})]\;,\label{hetra}\ee
where $H(S^{(A)},S^{(B)})$ is the quantity $$ H(S^{(A)},S^{(B)})=
 {\rm Tr_{\alpha}}\! \left[
U^{(AB)}\left(\Sigma^{\rm (pw)}_{-1} -S^{(A)}\Sigma^{\rm
(pw)}_{-1}S^{(A)\dagger}\right. \right.
$$
$$
\left.+S^{(A)}\Sigma^{\rm (ew)}_{-1}-\Sigma^{\rm
(ew)}_{-1}S^{(A)\dagger}\right) U^{(AB)\dagger}\,\left(
\Sigma^{\rm (pw)}_{1}\right.$$ \be \left.\left. - S^{(B)\dagger}
\Sigma^{\rm (pw)}_{1}S^{(B)}- \Sigma^{\rm (ew)}_{1}
S^{(B)}+S^{(B)\dagger} \Sigma^{\rm (ew)}_{1} \right)\right]
\;.\label{hqua}\ee By a lengthy computation, it is possible to
verify that the quantity $H(S^{(1)},S^{(2)})$ is symmetric under
the exchange of $S^{(1)}$ and $S^{(2)}$: \be
H(S^{(1)},S^{(2)})=H(S^{(2)},S^{(1)})\;.\ee By virtue of this
identity, the above formula for the power of heat transfer can be
rewritten as: \be W =\hbar \sum_{\omega
>0}\omega [n(\omega,T_1)-n(\omega,T_2)]H(S^{(1)},S^{(2)}) \;.\label{hetrabis}\ee
We stress once again that this formula provides an {\it exact}
expression for $W$ in terms of the scattering matrices of the
surfaces. We can  consider the simple special case of two planar
slabs. When the scattering matrices for two planar surfaces, given
in Eq. (\ref{plane}), are substituted into Eq. (\ref{hetrabis}),
the expression for the power of heat transfer takes the following
simple form:
$$W={\cal A} \times {\hbar}\sum_{\omega
>0} \omega [n(\omega,T_1)-n(\omega,T_2)]$$ $$\times
\sum_{\alpha}\left[\theta(k_z^2)
\frac{(1-|R_{\alpha}^{(1)}|^2)(1-|R_{\alpha}^{(2)}|^2)
}{|1-R_{\alpha}^{(1)}\,R_{\alpha}^{(2)}\,e^{2 i k_z
a}|^2}\right.$$ \be\left. + \,\theta(-k_z^2)\,4 \,e^{-2 a{\rm
Im}(k_z)}\,\frac{{\rm Im}(R_{\alpha}^{(1)}){\rm
Im}(R_{\alpha}^{(2)})
}{|1-R_{\alpha}^{(1)}\,R_{\alpha}^{(2)}\,e^{2 i k_z
a}|^2}\right]\;.\label{heatplane}\ee In the limit of large plates,
the above expression coincides with the known formula for the
power of heat transfer between two infinite plane-parallel
dielectric slabs separated by an empty gap \cite{volokitin}.

\section{Conclusions}

In conclusion, we have developed a new {\it exact} method for
computing Casimir forces and the power of heat transfer between
two plates of arbitrary compositions and shapes at different
temperatures, in vacuum. The method is based on a generalization
to systems out of thermal equilibrium of the the scattering
approach recently used to study the Casimir effect in non-planar
geometries \cite{emig,kenneth,genet}. Similarly to the equilibrium
case, we find that also out of thermal equilibrium the dependence
on shape and material appears only through the scattering matrices
of the intervening bodies. The expressions that have been obtained
are exact, and lend themselves to numerical or perturbative
computations once the scattering matrices for the desired geometry
are evaluated. Our results provide the tool for a systematic
investigation of the shape dependence of thermal proximity effects
in nanostructured surfaces, that could be of interest for future
applications to nanotechnology and to photonic crystals. In a
successive publication \cite{bimonte},  we shall use the formulae
derived in this paper to compute the Casimir force and the power
of heat transfer between two periodic dielectric gratings, like
those considered in last of Refs.\cite{genet}. The explicit form
of the scattering matrices for  rectangular gratings has been
worked out there, on the basis of a suitable generalization of the
Rayleigh expansion. At any finite order $N$ of the Rayleigh
expansion, the scattering matrices $S_{\alpha,\alpha'}$ are of the
form \be S_{\alpha,\alpha'} \equiv
\hat{S}(\tilde{{k}}_x,k_y)\,\delta({\tilde k}_x-{\tilde k}'_x)\,
\delta(k_y-k'_y)\;,\ee where ${\hat S}(\tilde{{k}}_x,k_y)$ is a
square matrix of dimension $2( 2N+1)$,  $\tilde{{k}}_x$ belongs to
the first Brillouin zone,  and $k_y$ is unrestricted. For
scattering matrices of this form, our explicit formulae for the
Casimir force and the power of heat transfer  can be evaluated
numerically quite easily, at least for sufficiently small $N$.\\

\noindent {\it Acknowledgements} The author thanks the ESF
Research Network CASIMIR for financial support.

%
%
%
%
%

\end{document}